\documentclass{article}
\usepackage{hyperref}
\usepackage{enumerate}
\usepackage{comment}
\usepackage{color}
\usepackage{graphicx}
\usepackage{float} 
\usepackage{multirow}
\newtheorem{thm}{Theorem}[section]
\newtheorem{hypothesis}[thm]{Hypothesis}

\begin{document}
\title{On the Role of Social Identity and Cohesion in Characterizing Online Social Communities}
\author{H. Purohit\footnote{Joint first authors.} \footnote{Kno.e.sis Center, Wright State University, \{hemant,amit\}@knoesis.org}, Y. Ruan\raisebox{4.5pt}{\footnotesize{$\ast$}}\footnote{Department of Computer Science and Engineering, The Ohio State University, \{ruan,fuhry,srini\}@cse.ohio-state.edu}, D. Fuhry\raisebox{4.5pt}{\footnotesize{$\ddagger$}}, S. Parthasarathy\raisebox{4.5pt}{\footnotesize{$\ddagger$}}, A. Sheth\raisebox{4.5pt}{\footnotesize{$\dagger$}}}

\maketitle
\begin{abstract}


Two prevailing theories for explaining social group or community structure are 
{\it cohesion} and {\it identity}. The social cohesion approach posits that social groups arise out of an aggregation of individuals that have mutual interpersonal attraction as they share common characteristics. These characteristics can range from common interests to kinship ties and from social values to ethnic backgrounds. In contrast, the social identity approach posits 
 that an individual is likely to join a group based on an intrinsic self-evaluation at a cognitive or perceptual level. In other words group members typically share an awareness of a common category membership.

In this work we seek to understand the role of these two contrasting theories in explaining the behavior and stability of social communities in Twitter. A specific focal point of our work is to understand the role of these theories in disparate contexts ranging from disaster response to socio-political activism. We extract social identity and social cohesion features-of-interest for large scale datasets of five real-world events and examine the effectiveness of such features in capturing behavioral characteristics and the stability of groups. We also propose a novel measure of social group sustainability based on the divergence in group discussion.
 Our main findings are: 1) Sharing of social identities (especially physical location) among group members has a positive impact on group sustainability, 2) Structural cohesion (represented by high group density and low average shortest path length) is a strong indicator of group sustainability, and 
3) Event characteristics play a role in shaping group sustainability,
 as social groups in transient events behave differently from groups in events that last longer.


\end{abstract}

\section{Introduction}
\label{sec:introduction}

Online social networks allow Internet users all over the globe to share information, exchange thoughts, and work collaboratively. All of those activities involve more than a single user, consequently, making questions on dynamics of \emph{online social groups} worthy of study. 
Especially, what factors influence an online social group's formation, its growth, and its sustainability? 

Prevalence of online social networks in the last decade has enabled computer scientists to answer questions of group sustainability and evaluate their solutions with large-scale 
experiments~\cite{backstrom2006group,shi2009user,kairam2012life,ruan2012prediction}. Despite the research progress made to date 
on community structure and group dynamics, there are at least three open questions to be answered:
\begin{itemize}
    \item 
    What is the relation between the findings of past research on group sustainability using structural characteristics and socio-psychological theories of group dynamics?    
    \item 
    How can existing theories on social group behavior guide us in identifying relevant features to model online social group sustainability?
    \item Online social group's sustainability not only depends on group size, but also on the divergence of its discussion content. How do we quantify this notion and what are the social group characteristics pertaining to it?
\end{itemize}

Over decades of study, social psychologists have proposed diverse explanations about the 
dynamics of a social group and its behavior. Two main frameworks among others are the \emph{social identity approach} and \emph{social cohesion approach}.

\noindent{\bf Social Identity Approach:}
\label{sec:social_identity_approach}
Social identity approach includes two closely related theories: \emph{social identity theory}~\cite{tajfel1971social} and \emph{self-categorization theory} \cite{turner1987rediscovering}.
In \cite{tajfel1971social}, Tajfel defines the concept of social identity as ``the individual's knowledge that he belongs to certain social groups together with some emotional and value significance to him of this group membership''. Therefore, group membership is the result of \textbf{``shared self-identification''} rather than ``cohesive interpersonal relationship'', and such shared identity leads to cohesiveness and uniformity, among other features~\cite{turner1982towards}. One commonly-cited evidence for social identity approach is team sports, where teammates are representing the same organization (a school, a club, or a country) and they are well aware to sustain the reputation of their associated identity. We refer to this approach as ``(social) identity'' in the following sections.

\noindent{\bf Social Cohesion Approach:}
\label{sec:social_cohesion_approach}
Social cohesion approach views social groups from a different perspective. Its hypothesis is that the necessary and sufficient condition for individuals to work as a group is the \textbf{cohesive social relationships between individuals}.
While social relationships exist for different reasons (e.g., kinship ties, or similar social values), we focus on a group's \emph{structural cohesion}, the collective result of those social relationships. 
Here, we adopt the definition by Lott and Lott~\cite{lott1965group} that interprets cohesiveness as \emph{mutual attraction} between individuals, which is slightly different from that used in~\cite{festinger1950spatial}. In accordance with this definition, the positive correlation between group cohesion and group's performance has been reported on various types of groups~\cite{mullen1994relation,beal2003cohesion}. We will denote this structural cohesion approach as ``(social) cohesion'' from now on. 

As noted above, social identity and social cohesion attribute group formation and sustainability to different factors. Identity approach posits that a social group is the result of members' collective awareness of some type of category membership.
In contrast, the conjecture of cohesion approach is that mutual attractions among individual pairs make them a group, implying that structural cohesion of member connections determines the sustainability of a social group. 


\label{sec:growth_rate_limitation}
To study the sustainability of social groups, a multitude of predictive models have been established to answer the question ``How many users will a social group have in the future?''. While group size and growth rate are intuitive measures, using them alone overlooks other important aspects in defining a social group's sustainability. One drawback, for example, is they do not capture the \textbf{stability of group membership}. Imagine that a group had five members previously, and later on four members left while nine new members joined in.
Although the group doubles in size, the low retention rate will have negative impact on its long-term sustainability.
Also, previous studies have not inspected the \textbf{divergence of content} generated in social groups. If each individual group member produces content of vastly different topics, it is harder for the community's voice to be heard.
Content coherence is especially critical for online discussion groups founded with a dedicated purpose (e.g. political rally~\cite{howard2012social}, disaster relief~\cite{sarcevic2012beacons}), and it is not captured by group size at all. Given the limitations of the simple measure of group size, alternative definitions of group sustainability are needed.

\noindent{\bf Main Results and Contribution:}
In this study, taking Twitter as our experimental platform, we quantify theoretic notions of social cohesion and social identity approaches from social science that accommodates to the characteristics of online social networks.
Social identity is computationally modeled via features of self-presentation in user profiles which could also encompass users' physical world identities. 
We represent social cohesion by structural features of the group's static friendship/follower network. 
These features incorporate guidance of the two theoretical approaches to capture users' social behavior from both physical and online world, and therefore, help us better understand the role of these theories in group behavior.

Furthermore, we propose a novel measure of social group sustainability, \emph{topic divergence}, based on the divergence of each individual member's discussion from the group's main line of discussion. Our two main hypotheses regarding group sustainability are:
\begin{hypothesis}
    The more 
    structurally
     cohesive a social group is, the lower topic divergence the group has.
    \label{thm:cohesion_hypothesis}
\end{hypothesis}

\begin{hypothesis}
    The more similar in identities a social group's members are, the lower topic divergence the group has.
    \label{thm:identity_hypothesis}
\end{hypothesis}

From experiments on five real-world datasets, we observe that 
1) Sharing of social identities (especially regional identity) among group members has a positive impact on group sustainability, 2) Structural cohesion (represented by high group density and low average shortest path length) is a strong indicator of group sustainability,
3) At least on Twitter, features based on the assumption of uni-directional interpersonal attraction have statistically equal explanatory power as features based on the assumption of mutual attraction, 
and 
4) Event characteristics affect online social group sustainability. Notably, during transient events like disasters, structurally cohesive social groups are less likely to exist, therefore, social identity of users can be utilized to create stable groups for help in the relief efforts. 

\section{Related Work}

Social network analysis has received greater attention in the last decade as online social networks have been evolving faster than ever.
Most of the studies took the path of network- or structure-centric approach to model community dynamics, aligning with more of the social cohesion approach. We discuss here some of the noteworthy studies covering different forms of group dynamics studied in the past. 

In the efforts to understand network structures of social networks at large scale, Mislove et al. \cite{mislove2007measurement} presented a study of Flickr, YouTube, LiveJournal, and Orkut networks. Their results confirmed the power-law, small-world, and scale-free properties of online social networks and observed that those networks contain a densely connected core of high-degree nodes; and that this core links small groups of strongly clustered, low-degree nodes at the fringes of the network. For community structures, Leskovec et al. \cite{leskovec2008statistical} studied the clustering problem on a wide range of real-world large networks and concluded that the ideal size for most community-like clusters was around 100 nodes. Kwak et al. \cite{kwak2010twitter} studied Twitter and presented various statistics for the entire Twittersphere, while reporting findings of a non-power-law follower distribution, a short effective diameter, and low reciprocity, and 4 degrees of separation in Twitter's follower network, differing from other human social networks. 

Following the structure-centric approach, link prediction and group formation problems were studied by various researchers. Notably, Liben-Nowell and Kleinberg \cite{liben2007link} surveyed various unsupervised methods on the link prediction problem and conducted extensive experiments on co-authorship networks.
Backstrom et al. \cite{backstrom2006group} proposed a model for network membership, growth and evolution by analyzing DBLP and LiveJournal social networks. They found that how individuals join communities and how communities grow depended on the underlying network structure, which supports structural cohesion in our discussion.
Taking a different path of a user-centric approach, Shi et al. \cite{shi2009user} studied the user behavior of joining communities on online forums. Among other features, authors studied the similarity between users and the similarity's relation with community overlap. 
Their results suggested that user similarity defined by frequency of communication or number of common friends was inadequate to predict grouping behavior, but adding node/user-level features could improve the fit of the model.

Among other notable efforts on group sustainability, Kairam et al. \cite{kairam2012life} analyzed long term (two years) dynamics of communities and modeled future community growth rate as a function of past growth or current size and age of the community. The study predicts growth rate and sustainability 
of the community and it was found that growth rate is correlated with current size and age of a group. For community sustainability, the size of the largest clique is the best feature.


In contrast to group-level studies, some researchers focused on user-level studies and therefore, efforts were made to understand user demographic on social networks. A noteworthy study by Rao et al. \cite{rao2011hierarchical} presented an approach for automatic creation of ethnic profiling of users, focusing on names as the key force. Building on the previous study, Pennacchiotti et al. \cite{pennacchiotti2011machine} proposed a machine learning approach to user classification on Twitter by analyzing user's friends, user posts and profile information.

In all of the discussion aforementioned, researchers modeled the group sustainability problem by either structural properties such as group size, or by evolution of volume in the content and activity. In our study, we present a systematic theoretical underpinning for group behavior by modeling the identity and cohesion phenomena into features-of-interest that cover not only structural- and activity-centric features studied in the past, but also user's identity-level characteristics. Furthermore, we propose measures to enable fine-grained understanding of group sustainability via content divergence, overcoming loopholes of size and growth rate based measures as discussed in Section~\ref{sec:introduction}. 

\section{Modeling and Experiments}
\label{sec:experiment}
Our experiment involves three major steps: 1) Identifying social groups, 2) Computing social identity and cohesion characteristics of users in the groups, and 3) Tracking the sustainability of the groups. Therefore, we first describe our data collection and social group identification approach, followed by quantitative modeling of each of the phenomena - social identity, social cohesion and group sustainability, necessary for experimentation of proposed research hypotheses in Section \ref{sec:introduction}.  
  
\subsection{Data Collection}
\label{sec:data_collection}
The Twitter \emph{Streaming} API provides real-time tweet collection. Alternatively, the Twitter \emph{Search} API provides keyword based search query, returning the 1500 most recent tweets in one response and excluding tweets from users who opt for privacy. To study the community forming around topic discussions for a specific event (denoted as ``event-oriented community''), we created a Streaming API based crawler that collected on-going tweet stream relevant to the event based on a seed keyword set, similar to \cite{ruan2012prediction}. For a keyword k, we crawl all tweets that mention k, K, \#k and \#K. The seed list of keywords and hashtags is kept up-to-date by first automatically collecting other hashtags and keywords that frequently appear in the crawled tweets and then manually selecting highly unambiguous hashtags and keywords from this list. We avoid the query drift problem by placing a human in the loop to ensure that ambiguous keywords are not crawled outside of context but only in combination with a contextually relevant keyword. One can also utilize a sophisticated computation method, such as Continuous Semantics framework~\cite{sheth2010continuous} to model the evolving knowledge and for finding highly relevant keywords for an event, but that is not the focus in this paper.

We also store associated metadata with the crawled tweets and for tweet posters, such as author location, followers and followees counts, description about the tweet poster, etc. We also crawl the social graph (i.e. follower list) of tweet posters who are part of the event-oriented community. For those users who activated privacy setting, no information was crawled, and their tweets were discarded from the dataset. 

To enable temporal analysis and reasoning, tweets are grouped into slices according to their associated time-stamp. In this paper each time slice is one day. Table \ref{table:dataset} shows various statistics about the datasets, two of which are about natural disaster (Type ``D'').

\begin{table}[h]
    \centering
    \small \addtolength{\tabcolsep}{-4pt}
    {
        \begin{tabular}{|c||c|c|c|c|}
            \hline
            \textbf{Event Name} & \textbf{Type} & \textbf{Duration} & \#\textbf{Tweets} & \#\textbf{Users} \\ \hline
			Hurricane Irene & D & 08/24-09/19, 2011 & 183K & 77K \\ \hline
            Hurricane Sandy & D & 10/27-11/07, 2012 & 4.9M & 1.8M \\ \hline
            India Anti-Corruption & non-D & 11/05-12/02, 2011 & 100K & 21K \\ \hline
            Occupy Wall Street & non-D & 11/05-12/02, 2011 & 2.1M & 331K \\ \hline
            Anti-SOPA & non-D & 01/19-02/19, 2012 & 744K & 389K \\ \hline
        \end{tabular}
    }
    \caption{Twitter data statistics centered on diverse set of events (D = natural disaster event)}
    \label{table:dataset}    
\end{table}

\begin{table}[h]
	\centering    
 \small \addtolength{\tabcolsep}{-5pt}    
    {
       \begin{tabular}{|c|c|c|c|c|c|}
           \hline
                            & \textbf{Lasting} & \textbf{Transient} \\ \hline
            \textbf{Loose}     & \begin{tabular}[x]{@{}c@{}}Occupy Wall Street,\\ India Anti-Corruption\end{tabular} & \begin{tabular}[x]{@{}c@{}}Hurricane Sandy,\\ Hurricane Irene \end{tabular} \\ \hline
           \textbf{Compact} & & Anti-SOPA \\ \hline                
        \end{tabular}
    }
\caption{Event classification~\cite{purohit2011understanding} based on event characteristics }
    \label{table:event-classification}
\end{table}

Analogous to~\cite{purohit2011understanding}, we note that events possess varying characteristics on the dimensions of activity, social significance, participant types, etc. Therefore, we also show event-classification for our datasets in the Table \ref{table:event-classification}. Loose and Compact event features reflect the nature of participants in the community, for example, the Anti-SOPA event was mostly driven by technology enthusiasts, a compact user set, and thus, it is a Compact event. Lasting and Transient features define the existence of vibe about the event, for example Occupy Wall Street protesters were long discussed in the social media, while after a week of Hurricane Sandy, nobody cared much about it except the people involved in the rebuilding phase of the disaster. 
Also, Hurricane events can be thought of as unexpected while protest events as deterministic, due to their organized coordinated sub-events. On the other hand, the involvement of the population type can also be used to suggest global versus the local scope of the events, for example, Hurricane Irene being local due to local coordination actors vs. Anti-SOPA being global due to global coordination of actors.  
Such event characterization will help us to diagnose the effects of event characteristics on the performance of social identity and cohesion to explain group sustainability.

\subsection{Identifying Social Groups}
Given all users in an event-oriented community, it is necessary to identify appropriate social groups on which quantitative analyses will be performed. Resultant social groups should reflect online interaction among users that is beyond simply using the same word in their tweets. Moreover, grouping criterion needs to be independent of any feature of social cohesion and social identity (defined in the following sections) so that the results are not biased.

To that end, we propose an approach of clustering users based on their interactions, which can be either retweet, reply or mention. A graph is created to represent those relationships, where vertices stand for users and edges indicate at least one interaction between two users during the whole dataset duration. We use a multi-level graph clustering algorithm \cite{satuluri2009scalable} to identify social groups, and remove groups that contain fewer than 10 members. We also remove groups that were active (i.e. at least one member posted a relevant tweet) in fewer than five time slices.
Clustering parameters are tuned such that the average size of the resultant groups is around 100, an empirical size of compact communities as observed in \cite{leskovec2008statistical}. Table \ref{tab:social_group_size} summarizes the information of each dataset's social groups.

\begin{table}[h]
    \centering
    \small 
    {
        \begin{tabular}{|c||c|c|c|}
            \hline
            & \# \textbf{Groups} & \# \textbf{Users} & \# \textbf{Users/Group} \\ \hline
            Hurricane Irene & 228 & 21,615 & 94.80 \\ \hline
            Hurricane Sandy & 3,438 & 340,401 & 99.01 \\ \hline
            India Anti-Corruption & 107 & 11,899 & 111.21 \\ \hline
            Occupy Wall Street & 2,549 & 239,927 & 94.13 \\ \hline
            Anti-SOPA & 1,389 & 149,490 & 107.62 \\ \hline
        \end{tabular}
    }
    \caption{Information of social groups identified from each event-oriented community}
    \label{tab:social_group_size}
\end{table}


\subsection{Quantifying Social Cohesion}
\label{sec:cohesion_stat}
To study the structural cohesion of social groups in a quantitative manner, 
we extract information from Twitter users' follower-followee graph. For each social group, we construct its corresponding node-induced sub-graph from the follower graph. Unlike many other online social network services, the follower relation on Twitter is directional, leading to three options when inducing the sub-graph:
\begin{itemize}
    \item \emph{Reciprocal}: \\ 
     An undirected edge will be created between two users only when both of them are following each other. This choice directly reflects the assumption of mutual interpersonal attraction in social cohesion approach. Statistics include density, transitivity (i.e. global clustering coefficient), average local clustering coefficient, and maximum average length of pairwise shortest path over all connected components (short-named ``average shortest path length'').
    \item \emph{Undirected}: \\
    An undirected edge will be created between two users if either of them is following the other. The underlying assumption is that a one-way interpersonal attraction is sufficient to keep the social group sustaining. Same group of statistics as in the reciprocal sub-graph are computed.
    \item \emph{Directed}:\\
        We also computed density and transitivity on the directed sub-graph for each social group, without converting to a undirected graph.
\end{itemize}

We are especially interested in the comparison of cohesion statistics calculated according to the reciprocal approach vs. the undirected approach. While both types of cohesion statistics reflect structural properties of social groups, the former encodes the condition of mutual attraction. From the perspective of social cohesion approach, the following hypothesis holds true:
\begin{hypothesis}
    Cohesion statistics of the reciprocal follower network are a better indicator of social group sustainability than that of 
    the undirected follower network.
    \label{thm:mutual_attraction_hypothesis}
\end{hypothesis}

The range for all cohesion statistics is $[0, 1]$, except for the average shortest path length as shown in Table \ref{tab:statistic_mean_std}. We report observations on the statistics in Section \ref{sec:discussion}.
We also notice the usage of structural cohesion's namesake in existing sociology literature \cite{moody2003structural,white2001cohesiveness}, where it was defined as the minimum number of nodes one need to remove to disconnect a graph.
We do not include this statistic as we find that almost all (more than 97\% of total) social groups contain at least one fringe node (whose degree is one) or singleton, meaning the value of this statistic for most of the groups will be at most one.

\subsection{Quantifying Social Identity}
\label{sec:identity_stat}
\begin{figure}[h]
  \centerline{ \includegraphics[scale=0.30]{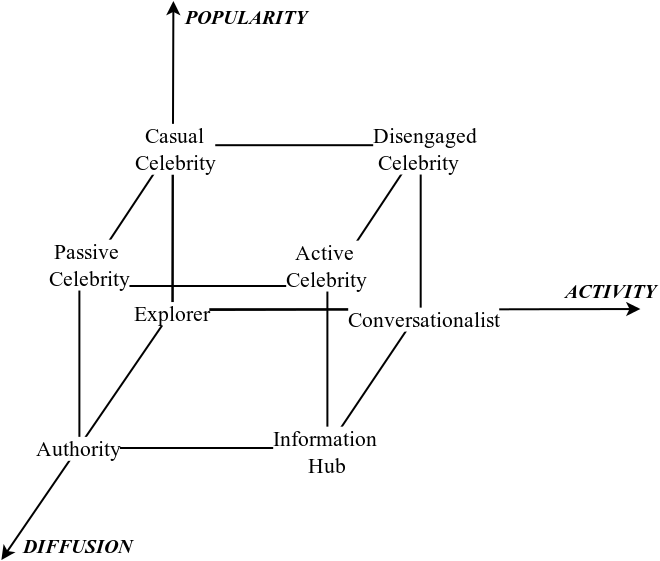} }
  \caption{AID Identity for users based on three action metrics } 
\label{fig:user-classification}    
\end{figure}

To quantify the social identity phenomenon, we extract identity features from the user profile information as well as activity, as we note that social behavior tends to associate the user with established identities via self-representation and with incentive-based identity via user actions (e.g., `active celebrity on Twitter'). For instance, people from New York like to be called `New Yorker', similarly University of Michigan students present themselves with the identity of the institution as `UMichigan' and computer engineers love to be called as `hackers' or `geeks'. We observe that there are various types of identity that we live with in our daily lives, ranging from regional, occupational expertise, organizational to cultural and religious identities, etc. We present our study covering some of these types in this paper.
From profile information, we can use location and interests metadata to extract the following types of social identities and for each such identity, we compute the entropy of its distribution in every social group:  
       \begin{itemize}
          \item \emph{Regional Identity}: \\ 
				Based on the `location' field of the user in the Twitter profile metadata, we map users to various geographical regions which tends to make an identity in our daily lives, e.g., `Indian' for an India based user, `Brit' for a UK based user and `New Yorker' for a New York based user. We choose to create state level and nation level identity of users in our study. Specifically, for an event, in the nation it occurs in, we map users belonging to the event's nation to the corresponding identity of states of that nation, while remaining users get a mapping of their respective national identities. We use Geonames dataset on Linked Open Data (LOD) and Google Maps API to convert user profile locations into latitude-longitude pairs as well as state and country level information. We note that this simple model of two regional levels (state and country) for identity can also be ported to a smaller scale (county and its next super-class, state) if an event is very specific to local interest. 
           \item \emph{Expertise Identity}: \\
           		Using `description' metadata in the user profiles, we map users to occupation and interests by entity spotting, which are also very common identities used in our daily lives, e.g., `Researcher' or `Artist' or `NFL player'. We fetch occupation titles using knowledge base sources, such as Wikipedia and the US department of Labor Statistics reports. We extend this knowledge base by human in the loop, because new conventions of social media have given rise to new forms of occupational interests (e.g., `blogger in digital marketing') which are not present in the formal 
occupation knowledge bases. At last, we classify occupation interests into 10 broader classes and thus give class labels to users, inspired by the domain classification on the news websites and also from the higher levels of occupation classes in the knowledge bases: \\  
           		\emph{ ACADEMICS, BUSINESS, POLITICS, TECHNOLOGY, BLOGGING, JOURNALISM, ART, SPORTS, MEDICAL, OTHERS }       \\    		
          We note that there can be more advanced methods to map user to expertise classes, but that is not our focus and we plan to keep exploration of more sophisticated methods for future work.          
    		\end{itemize}	             
Recent emergence in the services like \emph{Klout} or \emph{Foursquare} has brought a new convention of identity into our social lives where we participate in associating ourselves with incentive based identities, e.g., `Celebrity' by Klout, on Foursquare as `Mayor of Pier-39' for a popular San Francisco spot Pier-39. Therefore, in order to evaluate the effect of such identities derived from user actions in the social networks, we propose the following identity type based on the expertise presentation work of Purohit et. al~\cite{purohit2012user} and influence and passivity work of Romero et. al~\cite{romero2010influence}:
		\begin{itemize}	   
	   \item \emph{ Activity-Influence-Diffusion (AID) Identity}: \\
           Based on user actions on the platform (Twitter here), we use three metrics that contribute for building a user's AID identity: activity, popularity and diffusion strength. We model the activity metric by number of posts of the user, popularity metric by number of mentions of the user and diffusion strength by number of retweets of the user's posts. We compute scores on each of the three metric dimensions and then consider the 50th percentile threshold to create two levels on each of the dimensions, giving rise to 8 classes as shown in the Figure \ref{fig:user-classification}.
		\end{itemize}          		   
                In contrast with regional and expertise identities which are meaningful in the physical world, AID identity is a virtual world identity exclusively defined in the cyber realm. From our knowledge, few attempts have been made to study the impact of both online and offline identities on social networks.

The range of identity statistics is from 0 to $\ln(C)$, where $C$ is the number of unique classes in an identity type. In Table \ref{tab:statistic_mean_std} we summarize the basic information of each cohesion and identity statistic and report observations in Section ~\ref{sec:discussion}. The upper bounds of identity entropy values are included in brackets.

\begin{table}
    \small
    {
        \hspace{-1.5in}
        \begin{tabular}{|c||c|c|c|c|c|}
            \hline
                                                                                                  & Hurricane Irene        & Hurricane Sandy        & India Anti-Corruption  & Occupy Wall Street     & Anti-SOPA \\ \hline
            \multicolumn{6}{|l|}{\textbf{Structural Cohesion Statistics - Directed}} \\ \hline
            Density                                                                               & $0.03 \pm 0.05$        & $0.04 \pm 0.07$        & $0.01 \pm 0.01$        & $0.02 \pm 0.02$        & $0.03 \pm 0.06$ \\ \hline
            Transitivity                                                                          & $0.22 \pm 0.18$        & $0.23 \pm 0.21$        & $0.20 \pm 0.22$        & $0.23 \pm 0.19$        & $0.20 \pm 0.22$ \\ \hline
            \multicolumn{6}{|l|}{\textbf{Structural Cohesion Statistics - Reciprocal}} \\ \hline
            Density                                                                               & $0.02 \pm 0.04$        & $0.03 \pm 0.06$        & $0.00 \pm 0.01$        & $0.01 \pm 0.02$        & $0.02 \pm 0.05$ \\ \hline
            Transitivity                                                                          & $0.17 \pm 0.19$        & $0.22 \pm 0.24$        & $0.14 \pm 0.23$        & $0.19 \pm 0.22$        & $0.18 \pm 0.25$ \\ \hline
            Avg. Clustering Coef.                                                                 & $0.06 \pm 0.08$        & $0.08 \pm 0.11$        & $0.02 \pm 0.04$        & $0.05 \pm 0.06$        & $0.05 \pm 0.09$ \\ \hline
            Avg. Shortest Path Length                                                             & $2.24 \pm 1.30$        & $2.16 \pm 1.34$        & $1.71 \pm 0.99$        & $1.98 \pm 0.72$        & $1.75 \pm 0.97$ \\ \hline
            \multicolumn{6}{|l|}{\textbf{Structural Cohesion Statistics - Undirected}} \\ \hline
            Density                                                                               & $0.04 \pm 0.06$        & $0.05 \pm 0.08$        & $0.02 \pm 0.02$        & $0.02 \pm 0.03$        & $0.04 \pm 0.07$ \\ \hline
            Transitivity                                                                          & $0.16 \pm 0.17$        & $0.20 \pm 0.20$        & $0.13 \pm 0.17$        & $0.18 \pm 0.17$        & $0.19 \pm 0.20$ \\ \hline
            Avg. Clustering Coef.                                                                 & $0.13 \pm 0.12$        & $0.13 \pm 0.14$        & $0.07 \pm 0.09$        & $0.09 \pm 0.09$        & $0.11 \pm 0.13$ \\ \hline
            Avg. Shortest Path Length                                                             & $2.74 \pm 0.94$        & $2.74 \pm 1.38$        & $2.53 \pm 0.81$        & $2.65 \pm 0.83$        & $2.38 \pm 1.03$ \\ \hline
            \multicolumn{6}{|l|}{\textbf{Identity Statistics}} \\ \hline
            Regional Entropy                                                                      & $2.60 \pm 0.69 (5.28)$ & $2.50 \pm 0.74 (5.74)$ & $2.44 \pm 0.26 (4.94)$ & $2.75 \pm 0.51 (5.65)$ & $2.23 \pm 0.81 (5.53)$ \\ \hline
            Expertise Entropy                                                                     & $1.80 \pm 0.24 (2.30)$ & $1.14 \pm 0.43 (2.30)$ & $1.75 \pm 0.17 (2.30)$ & $1.67 \pm 0.19 (2.30)$ & $1.51 \pm 0.36 (2.30)$ \\ \hline
            AID Entropy                                                                           & $0.92 \pm 0.23 (2.08)$ & $0.99 \pm 0.21 (2.08)$ & $1.16 \pm 0.26 (2.08)$ & $1.18 \pm 0.24 (2.08)$ & $1.07 \pm 0.22 (2.08)$ \\ \hline
        \end{tabular}
     }
     \caption{Mean and standard deviation of structural cohesion/identity statistics. Identity entropy upper bounds are listed in brackets.}
     \label{tab:statistic_mean_std}
 \end{table}

\subsection{Measuring Social Group Sustainability}
\label{sec:sustainability_measure}
As discussed in Section \ref{sec:growth_rate_limitation}, there are limitations of using size and growth rate to measure the sustainability of a social group. Especially, growth rate will not capture the group's discussion divergence as well as its membership stability. Here, we introduce two alternative measures of social group sustainability, the first of which incorporates the notion of group discussion divergence and the second reflects membership stability.

\subsubsection{Topic Divergence}
To quantify the novel notion of discussion divergence within a group, we first construct a dynamic topic model \cite{blei2006dynamic} and infer the topics of discussion. Input into the topic model is a collection of vocabulary vectors, each of which represents event-related tweets posted by an author and is indexed by discrete time-stamps. The vocabulary includes words and phrases pertaining to the event (described in Section \ref{sec:data_collection}), as well as hashtags with the leading `\#' symbol stripped. The dynamic topic model has the advantage of modeling systematic topic shift (presumably due to event's progress) automatically, which allows us to investigate the true difference of an individual member's topic distribution to the corresponding group's topic distribution at any given time.

We let the number of topics $K$ be 3, and use default settings for other parameters for model inference\footnote{We used the implementation publicly available at \url{https://code.google.com/p/princeton-statistical-learning/downloads/detail?name=dtm_release.tgz}}. In Table \ref{tab:topic_model_top_words}, we list each topic's top vocabulary (excluding the event name itself) at three different stages of the event (beginning, middle and end)\footnote{For space constraint, we only show the lists of top words for Hurricane Sandy and Occupy Wall Street, the two largest datasets}. The transition of topic content is continual and smooth, and each topic is semantic distinct.

\begin{table}[h]
    \centering
    \small
    {
        \begin{tabular}{|c||c|c|c|}
            \hline
            \multicolumn{4}{|c|}{\textbf{Hurricane Sandy}} \\ \hline
            & \textbf{Beginning} & \textbf{Middle} & \textbf{End} \\ \hline
            \multirow{4}{*}{Topic 1} & tropical storm & red cross & red cross \\ \cline{2-4}
            & east coast & jersey shore & staten island \\ \cline{2-4}
            & canada & caused & mexico \\ \cline{2-4}
            & path & staten island & caused \\ \hline \hline
            \multirow{4}{*}{Topic 2} & new york & new york & new york \\ \cline{2-4}
            & state & new jersey & new jersey \\ \cline{2-4}
            & google & hurricane katrina & states \\ \cline{2-4}
            & android & media & hurricane katrina \\ \hline \hline
            \multirow{4}{*}{Topic 3} & frankenstorm & frankenstorm & frankenstorm \\ \cline{2-4}
            & halloween & fema & knicks \\ \cline{2-4}
            & east coast & halloween & fema \\ \cline{2-4}
            & atlantic & mitt romney & nyc \\ \hline \hline
            \multicolumn{4}{|c|}{\textbf{Occupy Wall Street}} \\ \hline
            & \textbf{Beginning} & \textbf{Middle} & \textbf{End} \\ \hline
            \multirow{4}{*}{Topic 1} & occupy & occupy & occupy \\ \cline{2-4}
            & protest & n17 & oo \\ \cline{2-4}
            & movement & nypd & occupyla \\ \cline{2-4}
            & occupytogether & brooklyn bridge & movement \\ \hline \hline
            \multirow{4}{*}{Topic 2} & movement & nypd & nypd \\ \cline{2-4}
            & us & movement & movement \\ \cline{2-4}
            & bahrain & protest & anonymous \\ \cline{2-4}
            & occupy movement & time & protest \\ \hline \hline
            \multirow{4}{*}{Topic 3} & occupy & occupy & p2 \\ \cline{2-4}
            & oo & p2 & tcot \\ \cline{2-4}
            & p2 & tcot & republican\\ \cline{2-4}
            & tcot & oo & teaparty \\ \hline
        \end{tabular}
        \caption{Top vocabulary of each topic at different event stages}
        \label{tab:topic_model_top_words}
    }
\end{table}

The inference process of the topic model returns a user's topic distribution at each time slice, denoted as $\beta_u^t$ for user $u$ at time $t$. Then we calculate the group topic distribution for group $g$ at time $t$ ($g_t$) as
\begin{equation}
    \beta_g^t(i) = \frac{\sum_{u \in g_t}{\beta_u^t(i)}}{|g_t|}, \forall i=1,2,3 ,
    \label{eqn:central_topic_distribution}
\end{equation}
and the \emph{topic divergence} of $g_t$ is defined as 
\begin{equation}
    TD(g_t) = \frac{\sum_{u \in g_t}{KL(\beta_g^t, \beta_u^t)}}{|g_t|} ,
    \label{eqn:topic_divergence}
\end{equation}
where $KL$ is the Kullback-Leibler divergence. Intuitively, this definition gauges the average divergence of each group member's topic distribution from the group's overall topic distribution. The greater the $TD$ value, the stronger indication of a group lacking conformity in discussion.

\subsubsection{Membership Stability}
The second sustainability measure we propose, called \emph{membership stability}, explicitly discounts a social group's size by its ``total change'' from the previous snapshot. For $g_t$, its \emph{membership stability} is defined as

\begin{equation}
    MS(g_t) = \frac{|g_t|}{|g_{t-1} \triangle g_t|+1} ,
    \label{eqn:membership_stability}
\end{equation}

where $\triangle$ is the set symmetric difference operator. The symmetric difference of group member sets at two sequential time slices is the set of users that left the group AND users that newly joined the group. This definition is inspired by a similar idea in \cite{asur2009event}, where the authors introduced the notion of \emph{stability index} to perform behavioral analysis of individuals in evolutionary graphs. 

\subsubsection{Growth Rate}
For comparison purposes, we also calculate the \emph{growth rate}, a widely-used size-based sustainability measure, for each $g_t$:
\begin{equation}
    GR(g_t) = \frac{|g_t|}{|g_{t-1}|} .
    \label{eqn:growth_rate}
\end{equation}

Table \ref{tab:sustainability_mean_std} provides an overview of sustainability measure's range for each event, where mean and standard deviation are calculated from each social group's average sustainability measure over time.
The values of topic divergence and growth rate spread more broadly, while the values of membership stability are more concentrated.
\begin{table}
    \centering
    \small
    {
        \begin{tabular}{|c||c|c|c|}
            \hline
                               & Divergence      & Stability       & Growth \\ \hline 
        Hurricane Irene        & $1.04 \pm 0.43$ & $0.53 \pm 0.10$ & $1.64 \pm 0.88$ \\ \hline
        Hurricane Sandy        & $0.71 \pm 0.47$ & $0.60 \pm 0.19$ & $1.66 \pm 0.99$ \\ \hline
        India Anti-Corruption  & $1.21 \pm 0.40$ & $0.67 \pm 0.12$ & $1.59 \pm 0.59$ \\ \hline
        Occupy Wall Street     & $1.34 \pm 0.38$ & $0.69 \pm 0.14$ & $2.17 \pm 1.70$ \\ \hline
        Anti-SOPA                   & $0.68 \pm 0.42$ & $0.51 \pm 0.21$ & $1.91 \pm 1.09$ \\ \hline
        \end{tabular}
    }
    \caption{Mean and standard deviation of sustainability measures}
    \label{tab:sustainability_mean_std}
\end{table}

\noindent{\bf{Correlation between Cohesion/Identity Statistics and Sustainability Measures:}}
We calculate the correlation coefficients between each social cohesion/identity statistic and each sustainability measure (topic divergence, membership stability, growth rate). We filter out social groups that contain fewer than ten members or have been active in fewer than five time slices. Each social group emits a tuple in the form of \emph{(cohesion/identity statistics, mean of sustainability measure over time)}. Tables \ref{tab:correlation_coef_divergence}, \ref{tab:correlation_coef_stability} and \ref{tab:correlation_coef_growth} summarize those values. Cells whose absolute value is greater than $0.25$ are boldfaced. We will analyze those results in details in the next section.

\begin{table}
    \small{
        \hspace{-1.3in}
        \begin{tabular}{|c||c|c|c|c|c|}
            \hline
                                                                                              & Hurricane Irene & Hurricane Sandy & India Anti-Corruption & Occupy Wall Street & Anti-SOPA \\ \hline
            \multicolumn{6}{|l|}{\textbf{Structural Cohesion Statistics - Directed}} \\ \hline
            Density                                                                           & \textbf{-0.33}  & \textbf{-0.33}  & -0.14                 & -0.11              & \textbf{-0.33} \\ \hline
            Transitivity                                                                      & 0.10            & 0.05            & 0.06                  & 0.16               & 0.07 \\ \hline
            \multicolumn{6}{|l|}{\textbf{Structural Cohesion Statistics - Reciprocal}} \\ \hline
            Density                                                                           & \textbf{-0.26}  & \textbf{-0.30}  & -0.11                 & -0.07              & \textbf{-0.27} \\ \hline
            Transitivity                                                                      & 0.15            & 0.06            & 0.24                  & 0.19               & 0.13 \\ \hline
            Avg. Clustering Coef.                                                             & 0.17            & -0.11           & \textbf{0.32}         & 0.16               & -0.01 \\ \hline
            Avg. Shortest Path Length                                                         & \textbf{0.57}   & 0.20            & 0.24                  & \textbf{0.43}      & \textbf{0.46} \\ \hline
            \multicolumn{6}{|l|}{\textbf{Structural Cohesion Statistics - Undirected}} \\ \hline
            Density                                                                           & \textbf{-0.35}  & \textbf{-0.34}  & -0.14                 & -0.13              & \textbf{-0.36} \\ \hline
            Transitivity                                                                      & 0.11            & 0.04            & 0.02                  & 0.23               & 0.11 \\ \hline
            Avg. Clustering Coef.                                                             & 0.22            & -0.09           & 0.05                  & 0.20               & 0.00 \\ \hline
            Avg. Shortest Path Length                                                         & \textbf{0.56}   & \textbf{0.28}   & \textbf{0.28}         & \textbf{0.37}      & \textbf{0.51} \\ \hline
            \multicolumn{6}{|l|}{\textbf{Identity Statistics}} \\ \hline
            Regional Entropy                                                                  & \textbf{0.43}   & \textbf{0.40}   & \textbf{0.28}         & \textbf{0.25}      & \textbf{0.58} \\ \hline
            Expertise Entropy                                                                 & \textbf{0.44}   & \textbf{0.64}   & \textbf{0.29}         & 0.18               & \textbf{0.39} \\ \hline
            AID Entropy                                                                  & \textbf{0.47}   & \textbf{0.28}   & 0.24                  & \textbf{0.58}      & \textbf{0.36} \\ \hline
        \end{tabular}
    }
    \caption{Correlation coefficients between structural cohesion/identity statistics and topic divergence}
    \label{tab:correlation_coef_divergence}
\end{table}
\begin{table}
    \small{
        \hspace{-1.3in}
        \begin{tabular}{|c||c|c|c|c|c|}
            \hline
                                                                                             & Hurricane Irene & Hurricane Sandy & India Anti-Corruption & Occupy Wall Street & Anti-SOPA \\ \hline
            \multicolumn{6}{|l|}{\textbf{Structural Cohesion Statistics - Directed}} \\ \hline
            Density                                                                          & 0.16            & 0.03            & 0.01                  & 0.10               & 0.08 \\ \hline
            Transitivity                                                                     & 0.13            & 0.06            & 0.01                  & 0.13               & 0.07 \\ \hline
            \multicolumn{6}{|l|}{\textbf{Structural Cohesion Statistics - Reciprocal}} \\ \hline
            Density                                                                          & 0.21            & 0.01            & -0.03                 & 0.12               & 0.02 \\ \hline
            Transitivity                                                                     & 0.18            & 0.03            & 0.06                  & 0.16               & 0.05 \\ \hline
            Avg. Clustering Coef.                                                            & \textbf{0.28}   & 0.02            & 0.10                  & 0.20               & 0.04 \\ \hline
            Avg. Shortest Path Length                                                        & 0.21            & 0.04            & 0.11                  & \textbf{0.28}      & 0.06 \\ \hline
            \multicolumn{6}{|l|}{\textbf{Structural Cohesion Statistics - Undirected}} \\ \hline
            Density                                                                          & 0.12            & 0.05            & 0.02                  & 0.09               & 0.12 \\ \hline
            Transitivity                                                                     & 0.17            & 0.08            & 0.03                  & 0.22               & 0.08 \\ \hline
            Avg. Clustering Coef.                                                            & \textbf{0.25}   & 0.09            & 0.03                  & 0.22               & 0.16 \\ \hline
            Avg. Shortest Path Length                                                        & 0.16            & 0.05            & 0.15                  & 0.19               & 0.06 \\ \hline
            \multicolumn{6}{|l|}{\textbf{Identity Statistics}} \\ \hline
            Regional Entropy                                                                 & -0.01           & 0.02            & -0.03                 & -0.12              & -0.03 \\ \hline
            Expertise Entropy                                                                & 0.13            & 0.15            & -0.04                 & 0.02               & -0.04 \\ \hline
            AID Entropy                                                                 & \textbf{0.42}   & 0.07            & \textbf{0.47}         & \textbf{0.62}      & 0.04 \\ \hline
        \end{tabular}
    }
    \caption{Correlation coefficients between structural cohesion/identity statistics and membership stability}
    \label{tab:correlation_coef_stability}
\end{table}
\begin{table}
    \small{
        \hspace{-1.3in}
        \begin{tabular}{|c||c|c|c|c|c|}
            \hline
                                                                                                    & Hurricane Irene & Hurricane Sandy & India Anti-Corruption & Occupy Wall Street & Anti-SOPA \\ \hline
            \multicolumn{6}{|l|}{\textbf{Structural Cohesion Statistics - Directed}} \\ \hline
            Density                                                                                 & -0.08           & -0.01           & 0.03                  & -0.14              & -0.06 \\ \hline
            Transitivity                                                                            & -0.02           & -0.19           & 0.07                  & -0.10              & -0.07 \\ \hline
            \multicolumn{6}{|l|}{\textbf{Structural Cohesion Statistics - Reciprocal}} \\ \hline
            Density                                                                                 & -0.09           & -0.02           & -0.01                 & -0.13              & -0.05 \\ \hline
            Transitivity                                                                            & 0.04            & -0.18           & -0.04                 & -0.09              & -0.06 \\ \hline
            Avg. Clustering Coef.                                                                   & 0.03            & -0.17           & -0.13                 & -0.14              & -0.08 \\ \hline
            Avg. Shortest Path Length                                                               & 0.18            & -0.27           & -0.14                 & -0.18              & -0.10 \\ \hline
            \multicolumn{6}{|l|}{\textbf{Structural Cohesion Statistics - Undirected}} \\ \hline
            Density                                                                                 & -0.08           & 0.00            & 0.03                  & -0.13              & -0.06 \\ \hline
            Transitivity                                                                            & 0.01            & -0.20           & -0.06                 & -0.12              & -0.08 \\ \hline
            Avg. Clustering Coef.                                                                   & 0.00            & -0.18           & 0.00                  & -0.17              & -0.11 \\ \hline
            Avg. Shortest Path Length                                                               & 0.11            & -0.26           & -0.09                 & -0.06              & -0.13 \\ \hline
            \multicolumn{6}{|l|}{\textbf{Identity Statistics}} \\ \hline
            Regional Entropy                                                                        & 0.20            & -0.21           & -0.07                 & 0.15               & -0.02 \\ \hline
            Expertise Entropy                                                                       & 0.05            & -0.16           & 0.15                  & -0.02              & -0.11 \\ \hline
            AID Entropy                                                                        & 0.02            & \textbf{-0.43}  & \textbf{-0.50}        & \textbf{-0.43}     & \textbf{-0.28} \\ \hline
        \end{tabular}
    }
    \caption{Correlation coefficients between structural cohesion/identity statistics and growth rate}
    \label{tab:correlation_coef_growth}
\end{table}

\section{Discussion}
\label{sec:discussion}
In this section, we discuss the results from Section~\ref{sec:experiment} and their implications.

\subsection{Identity and Cohesion Statistics}
\label{sec:identity_stat_disc}
We identify several interesting trends in the results reported in the Table \ref{tab:statistic_mean_std}. First, in general the entropy numbers\footnote{Note, it is important to normalize these numbers against the maximum entropy possible for each case.}
are higher for the Occupy Wall Street and India Anti-Corruption events,
the two on-the-ground political rally events, possibly because the offline interactions heavily involved in those events are not captured by online social identity statistics. Such distinction is most pronounced when comparing AID identity 
entropies of those two events with respect to the other three events. The social
groups in these two events tend to revolve around opinion leaders who often
help direct and orchestrate the movement (such individuals likely will have high AID values). Therefore social groups formed in those events generally have more diverse AID identity composition, reflecting the presence of opinion leaders as well as followers in groups. Next on the list, after these two events, is the Anti-SOPA rally, where Internet celebrities also play a 
leading role in influencing the discussion.
Another finding from Table \ref{tab:statistic_mean_std} is that groups have great divergence in terms of their memberships from different regions. This may simply be a reflection of the times and the fact that online social networks are bringing people closer together and that four out of five events have had significant media attention (SOPA, the odd one out in terms of media attention, has the lowest regional entropy).
Finally, we note that most events have low density values and their distributions of transitivity and clustering coefficient are often skewed toward zero. Both suggest sparse follower/followee connection in most social groups.

\subsection{Validating Hypotheses}
To validate the two hypotheses introduced in Section \ref{sec:introduction}, we check if the signs of correlation coefficients in Table \ref{tab:correlation_coef_divergence} agree with the induction from the hypotheses as following:
\begin{itemize}
\item Hypothesis \ref{thm:cohesion_hypothesis} posits that a more cohesive social group has a lower topic divergence. Higher density, transitivity and clustering coefficient signify a more cohesive structure, as does the lower value of average shortest path length. Therefore, we find 1) group density's negative correlation with topic divergence as well as 2) the positive correlation between average shortest path length and topic divergence are consistent with our hypothesis, suggesting group density and average shortest path length as sustainable group characteristics. On the other hand, the positive correlation with topic divergence for transitivity and clustering coefficient are in contrast with our hypothesis, as one would expect the social group with higher transitivity to have lower topic divergence. We suspect this counter-evidence has to do with the lack of triangles in social groups, as 
 analyzed below.

\item In Hypothesis \ref{thm:identity_hypothesis}, it is stated that if members of a social group are similar in identities, then the group should have low topic divergence. As identity entropy rises when group members' identities become more evenly-distributed, the induction from this hypothesis is that the identity entropy has positive correlation with topic divergence. Our results agree with this induction, as all three identities (regional, expertise, and AID) have positive correlation with topic divergence, for all events.\\
\end{itemize}

\subsection{Correlation Strength with Topic Divergence}
\noindent{\bf Identity Statistics:}
We note in Table \ref{tab:correlation_coef_divergence} that social identity statistics (especially regional identity entropy and AID identity entropy) have moderate to high positive correlation with topic divergence, implying a positive effect of identity characteristics on sustainability of the groups, and this holds true for all events. For social groups with stronger regional concentration, in-group discussions tend to be more location-specific and consistent, leading to a smaller degree of member-wise topic divergence, compared with groups whose members' locations are more disperse. Similarly, the presence of users with similar expertise or interest domain in a social group tends to keep the scope of discussions more focused.
For AID identity, we note that it is reflective of user actions, thus, we suspect that for the sake of maintaining their incentive-based action identity by lesser change in their actions, users tend to maintain a pattern of focused topic discussions in the groups.

\noindent {\bf Cohesion Statistics:}
\label{sec:correlation_between_cohesion_and_divergence}
For structural cohesion statistics, we find that patterns of correlation with topic divergence can be categorized into different groups:
\begin{itemize}
    \item First of all, triangle-based characteristics (global and average local clustering coefficient) show weak correlation with topic divergence in general. Many social groups have low clustering coefficients (see Table ~\ref{tab:statistic_mean_std}) due to the lack of triangles in their follower networks, hence the weak correlation. For future work, we plan to alleviate this issue by performing graph symmetrization, which discovers hidden similarity between nodes by comparing their inlink and outlink structures \cite{satuluri2011symmetrizations}.

\item Secondly, density statistics have moderate correlation with topic divergence for Hurricane Irene, Hurricane Sandy, and the Anti-SOPA rally, indicating that a better-connected social group tends to have a more cohesive discussion.\\
Why is this not the case for datasets of Occupy Wall Street and India anti-corruption movements?  As mentioned in Section \ref{sec:identity_stat_disc}, both of them are long-lasting events accompanied by an arguably more engaged offline component, whose information are not captured in cohesion statistics. Therefore, the density of online social groups is low (see Table \ref{tab:statistic_mean_std}), making it less indicative of sustainability for those two events.

\item Finally, average shortest path length shows consistency in its positive correlation with topic divergence. Similar to other cohesion statistics, the average shortest path length reflects the ``tightness'' of a social group. Compared with others, average shortest path length shows clearer dispersion in value, making the results of correlation analysis more meaningful.
\end{itemize}

\subsection{Reciprocal vs. Undirected Cohesion}
As introduced in Section \ref{sec:social_cohesion_approach}, the necessary and sufficient condition of social group formation via cohesion approach is the mutual attraction among group members. In our quantitative analysis, this translates to structural cohesion of the reciprocal follower graph, where two group members are connected only if they follow each other. We also derive a set of undirected structural cohesion statistics correspondingly, where two users are connected as long as either one is following the other. Therefore, undirected cohesion statistics reflect a weaker assumption that uni-directional interpersonal attraction is sufficient for social group sustainability.

Is mutual attraction really necessary for structural cohesion, and thus for sustainability of social groups? That is, can we validate Hypothesis \ref{thm:mutual_attraction_hypothesis}?
Again, we turn to Table \ref{tab:correlation_coef_divergence} for the answer, and perform \emph{one-sided binomial test} on the relative strength of correlation between both sets of cohesion statistics and topic divergence. Our null hypothesis is as follows:

\emph{$H_0$: It is equally likely that the (correlation) coefficient between a reciprocal statistic and topic divergence has a higher or lower absolute value than that of the coefficient between the respective undirected statistic and topic divergence.}

Our alternative hypothesis, corresponding to Hypothesis \ref{thm:mutual_attraction_hypothesis}, is:

\emph{$H_a$: The probability that the (correlation) coefficient between a reciprocal statistic and topic divergence has a higher absolute value than that of the coefficient between the respective undirected statistic and topic divergence, is more than 0.5.}

The test hypotheses are analogous to the situation where one wants to determine if a coin is fair ($H_0$), or its head is heavier than tail ($H_a$). Out of 20 observations (4 statistics X 5 events), only 9 times does reciprocal statistic's coefficient have a higher absolute value, corresponding to a p-value of 0.7483. With such a large p-value, we cannot reject $H_0$ in favor of $H_a$, thus there is little evidence supporting Hypothesis \ref{thm:mutual_attraction_hypothesis}. Therefore our results suggest that mutual attraction is not a necessary condition of structural cohesion and group sustainability. Note that, however, this should not be interpreted as the opposite belief that undirected cohesion statistics are a better indicator of topic divergence than reciprocal cohesion statistics. The p-value in that case is 0.4119, which is not significant either.

\subsection{Correlation with Other Measures of Sustainability}
Moving to Tables \ref{tab:correlation_coef_stability} and \ref{tab:correlation_coef_growth}, we observe that none of the cohesion or identity statistics, except AID entropy, has a high correlation with either membership stability or growth rate across all datasets.
It supports our argument that size-based measures for community sustainability may not be sufficient and need to be complemented by content coherence-based measures for enhanced understanding of sustainability of social groups.

\subsection{Effects of Event Characteristics}
Tables  \ref{tab:correlation_coef_divergence},  \ref{tab:correlation_coef_stability}  and  \ref{tab:correlation_coef_growth} highlight interesting differences in the effectiveness of cohesion and identity approaches in modeling sustainability across various event types:
\begin{itemize}
\item Table \ref{tab:correlation_coef_divergence} shows that transient types of events (Hurricane Irene, Sandy and Anti-SOPA) have better correlation of topic divergence measure (sustainability metric) with features of social identities as compared to those of social cohesion. It is perhaps due to the fact that groups in such volatile events form in an ad-hoc setting, where groups are less likely to have existing cohesively connected users, undermining the effects of features corresponding to social cohesion here. Therefore, discussions can be highly dependent on the characteristics of participants of the group, their personal behavior and identities. 

\item It is interesting to note a high correlation pattern for Anti-SOPA as compared to Occupy Wall Street and India Anti-corruption protest events, for both social identity and cohesion measures in Table \ref{tab:correlation_coef_divergence}. It may be due to the nature of coordination, where one is a cyber protest, requiring better organization of activities online, thus more focused representation of activities (especially the one where major websites including Wikipedia had taken their content off, replacing it with black screen to protest), while Occupy Wall Street and India Anti-corruption are more ground-run protests and events were coordinated by physically meet-ups.
\end{itemize}

\section{Future Work}
 We plan to extend our measures of social identity and cohesion with more features, such as ethnic and religious social relationships which can enhance our analysis with more insights into how real-world groups unfold over time. We also plan to perform proposed analyses on other social networks, such as Facebook, LinkedIn and online forums, and on the co-authorship network of DBLP, to see if they show a similar social phenomena of group dynamics. We also plan to explore the usage of Twitter Lists subscriptions to create new forms of social cohesion and identity measures.

\section{Conclusion}

This study focuses on characterizing online social group sustainability by socio-psychological theories of group bonding and attachment - \emph{social identity} and \emph{social cohesion}. This study on Twitter is not only the first to quantify theoretic notions of identity and cohesion in the social groups, but also to present various approaches to model sustainability of the group beyond past approaches of structure-based properties such as group size. 
Features inspired by both theories are found to correlate with social group sustainability well. 
We also observe an effect of event characteristics on stability of the groups and report our observations by large scale experimentation on a diverse set of real-world events. 

\section{Acknowledgement}
We thank our colleagues for useful feedback and NSF for sponsoring SoCS Grant IIS-1111118 and IIS-1111182 titled as `Social Media Enhanced Organizational Sensemaking in Emergency Response'.

\bibliographystyle{abbrv}
\bibliography{www2013}
\end{document}